\documentclass[preprint,12pt,review]{elsarticle}

\usepackage{amssymb}
\usepackage{amsmath}

\journal{Physics Letters B}

\begin{document}

\begin{frontmatter}



\title{Hidden Glashow resonance in neutrino--nucleus collisions}

\author{I. Alikhanov\corref{cor1}}

\cortext[cor1]{{\it Email address: {\tt ialspbu@gmail.com}}}

\address{Institute for Nuclear Research of the Russian Academy of Sciences,
60-th October Anniversary pr. 7a, Moscow 117312, Russia\\
Research Institute for Applied Mathematics and Automation, Shortanova Str. 89-A, Nalchik 360000, Kabardino-Balkarian Republic, Russia}

\begin{abstract}
Today it is widely believed that $s$-channel excitation of an on-shell~$W$ boson, commonly known as the Glashow resonance, can be initiated in matter only by the electron antineutrino in the process $\bar\nu_ee^-\rightarrow W^-$  at the laboratory energy  around~6.3~PeV. In this Letter we argue that the Glashow resonance within the Standard Model also occurs in neutrino--nucleus collisions. The main conclusions are as follows.
1)~The Glashow resonance can be excited  by both~neutrinos and~antineutrinos of all the three flavors scattering in the Coulomb field of a nucleus.
2)~The Glashow resonance in a neutrino--nucleus reaction does not manifest itself as a Breit--Wigner-like peak in the cross section but the latter exhibits instead a slow logarithmic-law growth with the neutrino energy. The resonance turns thus out to be hidden. 
3)~More than~$98\%$ of~$W$  bosons produced in the sub-PeV region in neutrino-initiated reactions in water/ice  will be from the Glashow resonance. 4)~The vast majority of the Glashow resonance events in a neutrino detector is expected at energies from a few TeV to a few tens of TeV, being mostly initiated by the conventional atmospheric neutrinos dominant in this energy range. 
Calculations of the cross sections for Glashow resonance excitation  on the oxygen nucleus as well as on the proton are carried out in detail. 
The results of this Letter can be useful for studies of neutrino interactions at large volume water/ice neutrino detectors. For example, in the IceCube detector one can expect~0.3 Glashow resonance events  with  shower-like topologies and the deposited energies above~$300~\text{TeV}$ per year. It is therefore likely already to have at least one Glashow resonance event in the IceCube data set.
\end{abstract}


\end{frontmatter}

\section{Introduction}
A single resonance formed by two colliding particles manifests itself as a dramatic rise of the corresponding cross section to a peak over a relatively narrow range of the collision energy. Such a resonance, usually referred to as an $s$-channel resonance, serves as an intermediate state between the incident particles and the outgoing products of its subsequent decay. 
Formally, the resonant enhancement of the cross section takes place due to the pole-like behavior of the probability amplitude for this process $M\propto(s-m^2+im\Gamma)^{-1}$ ($s$~is the total center-of-mass energy squared of the colliding particles, $m$ and $\Gamma$ are the mass and the width of the resonance).
 As a result, in the vicinity of the pole $s=m^2$, the dependence of the cross section on the energy has the well known Breit--Wigner shape $\sigma(s)\propto\left((s-m^2)^2+m^2\Gamma^2\right)^{-1}$.
If $\Gamma\ll m$, the cross section may be approximated by the  Dirac delta-function so that $\sigma(s)\propto\delta(s-m^2)$.
This is the so-called narrow widths approximation which sometimes substantially reduces the complexity of scattering calculations.

It is difficult to overestimate the role played by the experimental observations of resonances of this kind in the development of elementary particle physics. Just recall the milestone discoveries of the $\Delta^{++}(1232)$ in $\pi^+p$ scattering~\cite{delta1,delta11,delta2}, $J/\Psi$ in $e^+e^-$ annihilation~\cite{jipsi} and the precise determination of the fundamental input parameters of the Standard Model by investigating the $Z^0$ peak at electron--positron colliders~\cite{pdg}.

Along with the impressive success of the Standard Model we have witnessed for decades, there are processes predicted by this model but the existence of which has yet to be proven by experiment. This is the case, for instance, in $\bar\nu_e$  scattering on electrons whose cross section should have a sharp resonance peak occasioned by $s$-channel excitation of the real $W^-$ boson~\cite{Glashow_res}, $\bar\nu_ee^-\rightarrow W^-$, commonly known as the Glashow resonance. The Glashow resonance can be effectively searched for in water/ice neutrino detectors~\cite{berezinsky} through the reaction $\bar\nu_ee^-\rightarrow W^-\rightarrow~\text{anything}$ initiated by cosmic-ray electron antineutrinos of energies of about $m_W^2/2m_e=6.3~\text{PeV}$ (1~PeV = $10^{15}~\text{eV}$). 
With the completion of the IceCube kilometer-scale neutrino telescope located at the South Pole~\cite{icecube_rev}, the idea of observing the Glashow resonance is again in the focus of attention of physicists~\cite{Glashow_res_cube1, Glashow_res_cube2, Glashow_res_cube3, Glashow_res_cube4, Glashow_res_cube5,Glashow_res_cube6}. Moreover, it has already been proposed to interpret the PeV cascade events ($\simeq1.04~\text{PeV}$, $\simeq1.14~\text{PeV}$, $\simeq2.00~\text{PeV}$) recently reported by the IceCube experiment~\cite{icecube_2events,icecube_28,icecube_new} in terms of the Glashow resonance~\cite{Glashow_res_cube7,Glashow_res_cube8}. Even though there is some probability that the Glashow resonance emerges in the interval between~1 and~6.3 PeV~\cite{below_pev1}, the relatively wide energy gap, $\gtrsim4~\text{PeV}$, still separates the observed events from the expected position of the resonance peak. Anyway, no convincing evidence for the existence of the Glashow resonance has been found  up until now and its discovery would undoubtedly be a crucial test of the Standard  Model.  
The Standard Model also predicts the same resonant scatterings for other lepton pairs, $\nu_ee^+\rightarrow W^+$, $\overset{\text{{\tiny (}}-\text{{\tiny )}}}{\nu_{\mu}}\mu^\mp\rightarrow W^\mp$, $\overset{\text{{\tiny (}}-\text{{\tiny )}}}{\nu_{\tau}}\tau^\mp\rightarrow W^\mp$, however the explicit presence of electrons in the target justifies the high theoretical and experimental attention that the channel $\bar\nu_ee^-\rightarrow W^-$ has received, as compared to the former ones. 

Meanwhile, there are processes in which although an $s$-channel resonance is formed, this will not be indicated by a peak in the cross section. This occurs, for example, when either of the two particles annihilating into the resonance has not a definite momentum but is characterized instead by a probability that it carries a certain momentum, like partons in the nucleon. Consider, in particular, $s$-channel single leptoquark production in neutrino--nucleon collisions through direct neutrino--quark fusion~\cite{Berezinsky:1985yw,lq_1997,Carena:1998gd,Anchordoqui:2006wc,me_lq,Barger}. The corresponding cross section within the narrow width approximation reads $\sigma_{LQ}(s)\propto\int dx\,q(x)\delta(x s-m^2_{LQ})=q\left(m^2_{LQ}/s\right)/s$~\cite{lq_1997} ($q(x)$ is the probability density that the nucleon will manifest itself as the relevant quark with fraction~$x$ of the initial nucleon momentum).
Since $q(x)\propto$~${1/x^{1+\lambda}}$, the cross section scales as~$\sigma_{LQ}(s)\propto s^{\lambda}$ (typically, $\lambda\sim0.3$)~\cite{me_lq}. The latter result clearly illustrates that though the leptoquarks are resonantly produced in the $s$-channel, the experimentally observable cross section does not exhibit the canonical resonance structure but just a slow monotonic power-law growth with the center-of-mass neutrino--nucleon collision energy. And it may happen that a researcher analyzing a similar process without knowing about the underlying resonance will stay unaware of its existence. In this sense such $s$-channel resonances turn out to be \textit{hidden}. 

A recent Standard Model analysis of neutrino--photon interactions strongly suggests that a resonant mechanism is also responsible for production of the on-shell $W$ bosons in reactions $\overset{\text{{\tiny (}}-\text{{\tiny )}}}{\nu_l}\gamma\rightarrow l^\pm W^\mp$ $(l=e,\mu,\tau)$~\cite{me_2015}.
Namely, the $W$~bosons are produced through the $s$-channel  $\overset{\text{{\tiny (}}-\text{{\tiny )}}}{\nu_l}l^\mp\rightarrow~W^\mp$ subprocesses, in other words, through the Glashow resonance. The incident charged leptons on which neutrinos resonantly annihilate emerge due to photon splitting $\gamma\rightarrow l^+l^-$.

It is remarkable that the reactions~$\overset{\text{{\tiny (}}-\text{{\tiny )}}}{\nu_l}\gamma\rightarrow l^\pm W^\mp$ make the observation of the Glashow resonance experimentally feasible for all the three lepton flavors of the Standard Model, $e$, $\mu$, $\tau$, by impinging high-energy neutrinos on nuclear targets or, more precisely, on the Weizs\"acker--Williams photons generated by the nuclei. At the same time, the required neutrino energies for these reactions to proceed are far below the PeV region in which the Glashow resonance is eagerly awaited today. 
For example, in collisions with protons and oxygen nuclei in water/ice,  $\overset{\text{{\tiny (}}-\text{{\tiny )}}}{\nu_l}p\rightarrow W^\mp X$,  $\overset{\text{{\tiny (}}-\text{{\tiny )}}}{\nu_l}\text{$^{16}$O}\rightarrow W^\mp X$,  the threshold neutrino energies are just~$\sim10$~TeV~\cite{me_2015}. Such thresholds allow to exploit even the conventional atmospheric neutrino flux to probe the Glashow resonance at neutrino telescopes. The dominant hadronic decay modes of the $W$ boson will give showers highly boosted along the direction of the incident neutrinos inside a detector.

The considered processes were studied for the first time in~\cite{seckel} where it was pointed out that the lepton propagators enhance the cross sections near the threshold.  In addition to the production in the Coulomb field of a nucleus, W bosons can also be singly produced by neutrinos in a magnetic field~\cite{kuznetsov}. These processes may have some implications for astrophysics and cosmology~\cite{seckel,me_other}.

In this Letter we calculate the cross sections for Glashow resonance excitation in $\overset{\text{{\tiny (}}-\text{{\tiny )}}}{\nu_l}p$ and $\overset{\text{{\tiny (}}-\text{{\tiny )}}}{\nu_l}\text{$^{16}$O}$ collisions within the equivalent photon (Weizs\"acker--Williams) approximation. We take into account: 1)~coherent neutrino scattering on the $^{16}\text{O}$ nucleus; 2)~elastic neutrino--proton scattering; 3)~inelastic neutrino--proton and  neutrino--neutron collisions. We discuss theoretical uncertainties of the calculations and  signatures of the hidden Glashow resonance in water/ice. We also evaluate the corresponding total event rates from the different components of the neutrino flux reaching ground level.

\section{Unveiling the Glashow resonance in $\nu\gamma$ interactions}

Let us first consider exclusive production of the on-shell~$W$ bosons in the following reactions:

\begin{equation}
\overset{\text{{\tiny (}}-\text{{\tiny )}}}{\nu_l}\gamma\rightarrow l^\pm W^\mp,\,\,\,(l=e,\mu,\tau).\label{intro_main}
\end{equation}

\begin{figure}
\includegraphics[width=1.\textwidth]{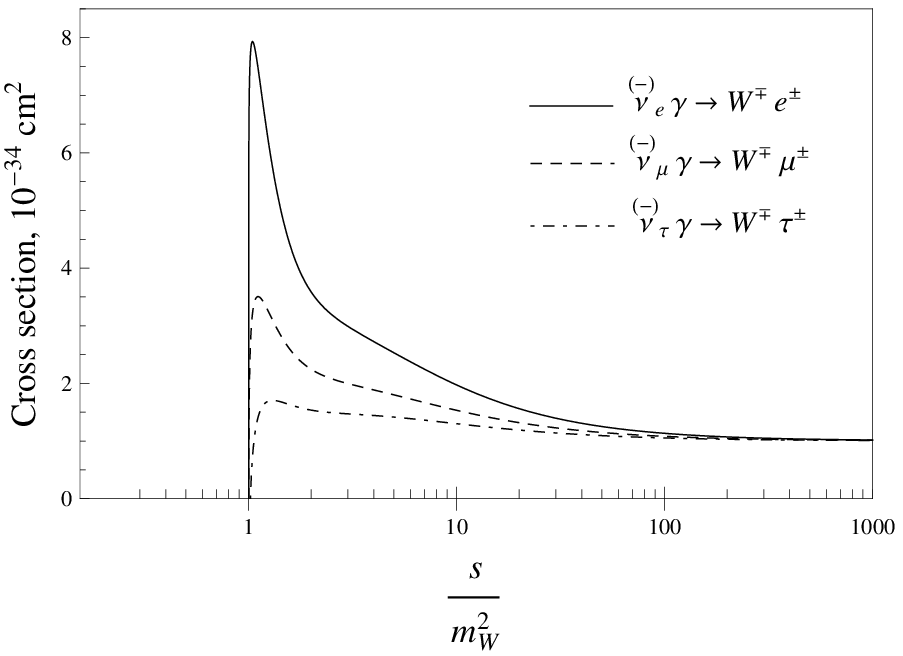}
\caption{}
\label{cross_sm}
\end{figure}

The corresponding leading order cross sections calculated within the Standard Model read~\cite{seckel,me_plb13}

\begin{equation}
\sigma_{l}=\sqrt{2}\alpha G_F\left[2(1 - \xi)(1 + 2\xi^2 +\xi^2\log{\xi}) 
   +\xi(1 - 2\xi + 2\xi^2)\log\left(\frac{m_W^2}{m_l^2}\frac{(1-\xi)^2}{\xi}\right)\right],\label{tree}
\end{equation}

where $\xi=m_W^2/s$, $m_l$ is the mass of the final lepton, $G_F$ is the Fermi
constant and $\alpha$ is the fine structure constant. Note that since CP is conserved here, there is no difference in~$\sigma_l$ for $\nu_l\gamma\rightarrow l^-W^+$ and $\bar\nu_l\gamma\rightarrow l^+W^-$. These cross sections are presented as functions of~$s/m_W^2$ in Fig.~\ref{cross_sm}.

\begin{figure}
\includegraphics[width=0.8\textwidth]{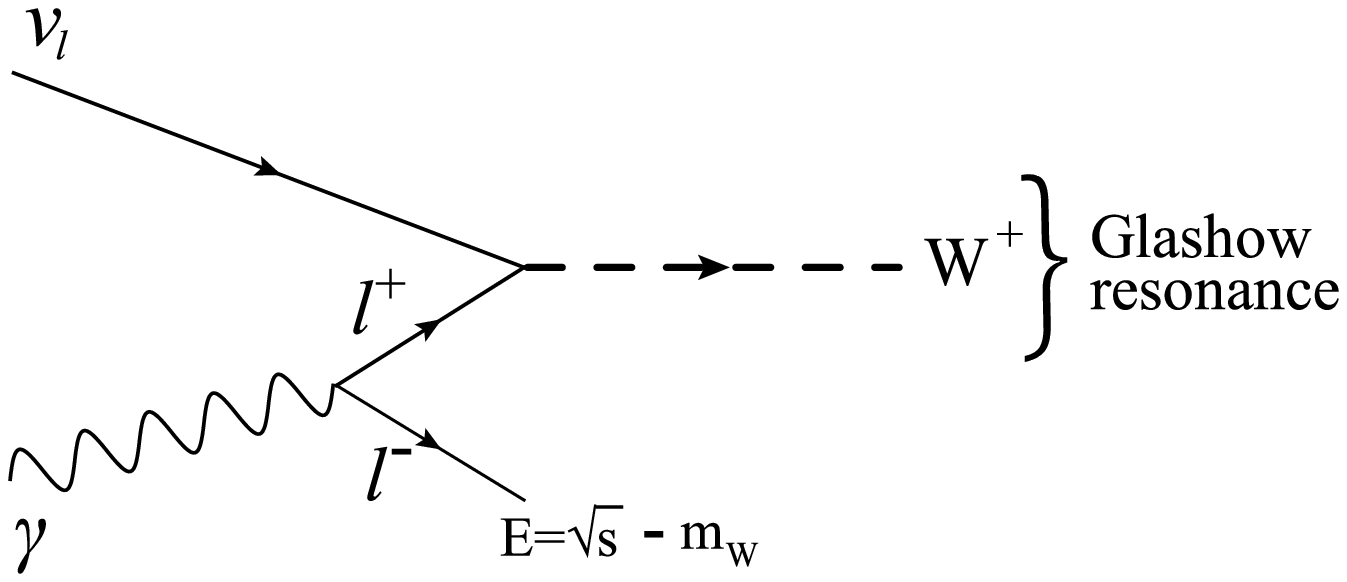}
\caption{}
\label{fig:s_chan}
\end{figure}

The Standard Model strongly suggests that the $W$ bosons in these reactions emerge through the Glashow resonance~\cite{me_2015}.  The underlying mechanism is schematically illustrated in Fig.~\ref{fig:s_chan} for the case of neutrinos (the same holds, of course, for antineutrinos).  The  ingoing neutrino resonantly annihilates on the positively charged lepton coming from photon splitting $\gamma\rightarrow l^+l^-$. Even if the $\nu_l\gamma$ collision energy, $\sqrt{s}$, exceeds the mass of~$W^+$, the outgoing~$l^-$ carries away the energy excess, $E=\sqrt{s}-m_W$,  and turns thus the~$\nu_ll^+$ pair to the resonance pole. This resembles the well known initial state radiation in $e^+e^-$~collisions when emission of photons from the initial electron (positron) before $e^+e^-$ annihilation essentially modifies the shape of a narrow resonance curve: the curve becomes wider, a suppression of the resonance maximum is observed and the so-called radiation tail appears to the right of the resonance pole~\cite{lipatov}. As seen from Fig.~\ref{cross_sm}, similar features are exhibited by the reactions~$\overset{\text{{\tiny (}}-\text{{\tiny )}}}{\nu_l}\gamma\rightarrow l^\pm W^\mp$: their cross sections resonantly grow at the pole~$s=m_W^2$   and then gradually decline, also forming tails due to emission of the charged leptons from the incident photon. Put another way, each of the cross sections in Fig.~\ref{cross_sm} represents in fact the Glashow resonance peak smeared out by the final charged lepton momentum and simultaneously suppressed due to (roughly) an extra vertex factor~$\alpha$ for the transition~$\gamma\rightarrow l^+l^-$~\cite{me_2015}. 

\section{Theoretical framework}
\begin{figure}
\includegraphics[width=1.\textwidth]{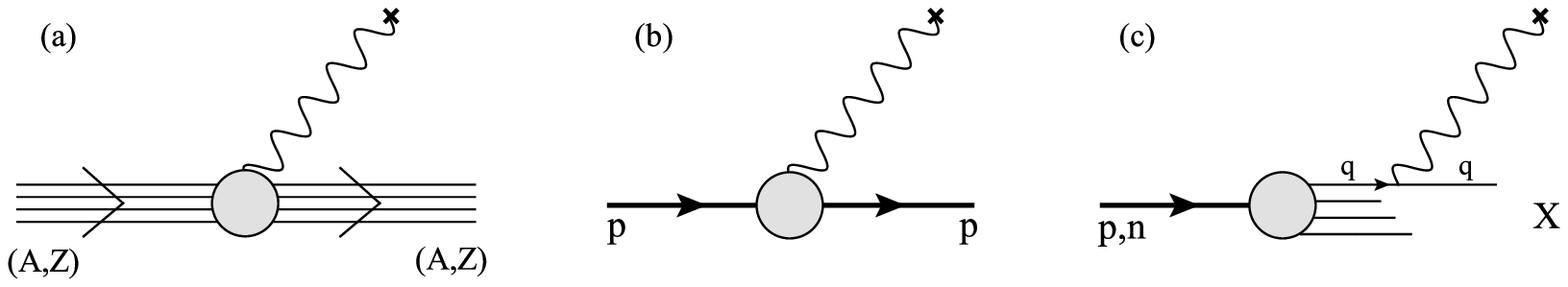}
\caption{}
\label{fig:contribs}
\end{figure}

Since the Glashow resonance appears, as discussed above, in $\nu\gamma$ interactions, it can therefore be excited in neutrino--nucleus collisions as well, when neutrinos interact with the equivalent (Weizs\"acker--Williams) photons of the nuclear target. Consider a nucleus composed of $Z$ protons and $A-Z$ neutrons. 
In practice, it is convenient to have the cross sections per nucleon, so that for $
\overset{\text{{\tiny (}}-\text{{\tiny )}}}{\nu_l}+(A,Z)\rightarrow (W^\mp)_{\text{Res}}+X$ one can write

\begin{equation}
\sigma_{Nl}(s)=\int dx\,\gamma(x)\sigma_l(xs),
\label{nucleus}
\end{equation}
where $\gamma(x)$ is the equivalent photon distribution, $\sigma_l(s)$ is given by~(\ref{tree}).
The function $\gamma(x)$ consists of four components corresponding to the four possibilities of interaction:

\begin{equation}
\gamma(x)=\frac{1}{A}\left[\gamma_{\text{coherent}}(x)+Z\gamma_{\text{p\,el}}(x)+Z\gamma_{\text{p\,inel}}(x)+(A-Z)\gamma_{\text{n\,inel}}(x)\right],
\label{eq:contribs}
\end{equation}

 namely, the coherent photon content of the nucleus when the latter radiates off a photon as a whole without break-up, $\gamma_{\text{coherent}}(x)$, the  elastic photon content of the proton when a separate proton of the nucleus emits the photon, $\gamma_{\text{p\,el}}(x)$,  the inelastic photon content of the nucleon (proton and neutron) when the photon comes from a separate nucleon which subsequently breaks up, $\gamma_{\text{p,n\,inel}}(x)$. These components are sketched in Fig.~\ref{fig:contribs}.  It should be noted that the limits of integration in~(\ref{nucleus}) depend on the mass of the object which radiates the photon. Thus for the integration over the coherent part, one has to take $x_{\text{min}}=m_W^2/2Am_NE_{\nu}$ and $x_{\text{max}}=\left(1-\sqrt{Am_N/2E_{\nu}}\right)^2$, where $m_N$ is the nucleon mass, $E_{\nu}$ is the neutrino energy in the laboratory reference frame.  For the integration over the remainder three components, the lower and upper limits are $x_{\text{min}}=m_W^2/2m_NE_{\nu}$,  $x_{\text{max}}=\left(1-\sqrt{m_N/2E_{\nu}}\right)^2$, respectively. The choice of these limits becomes obvious if to recall that $x$~is the fraction of the initial nucleus/nucleon energy carried away by the photon. Since $E_{\nu}\gg Am_N$, $x_{\text{max}}$ can in principle be set to unity, as it is often done in the parton model. 

\section{Numerical calculations}
\begin{figure}
\includegraphics[width=1.\textwidth]{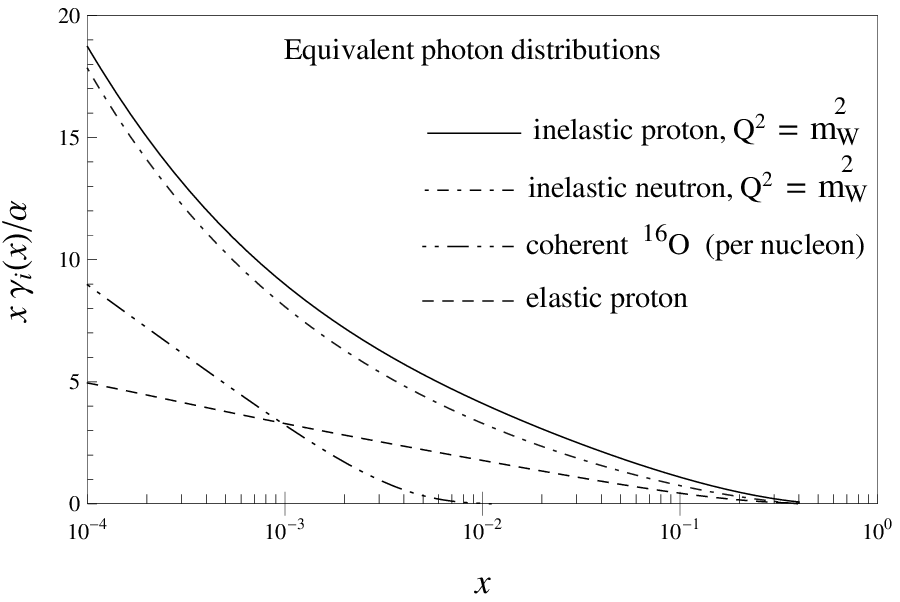}
\caption{}
\label{fig:pdf}
\end{figure}

\begin{figure}
\includegraphics[width=1.\textwidth]{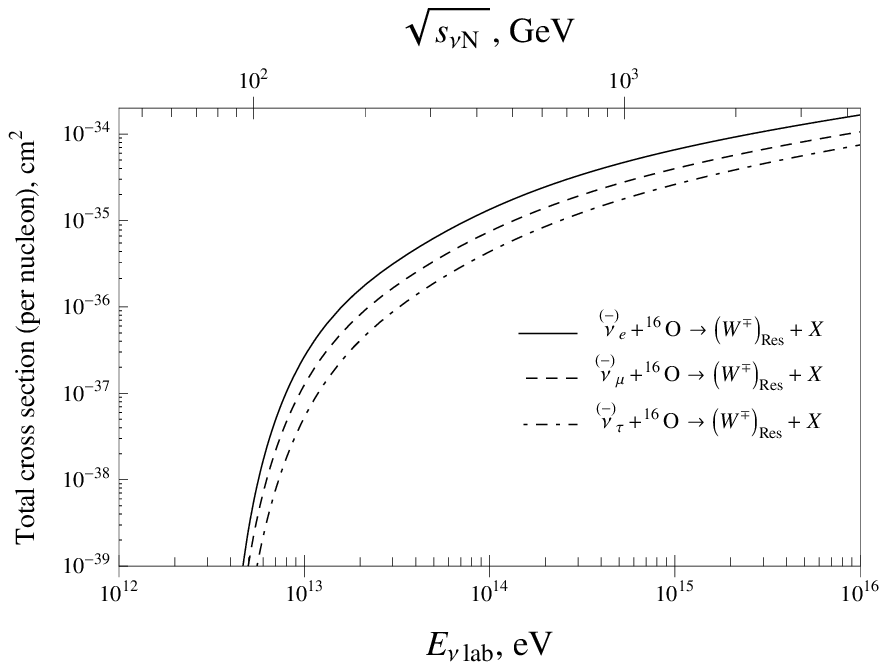}
\caption{}
\label{fig:cr_total}
\end{figure}

\begin{figure}
\includegraphics[width=1.\textwidth]{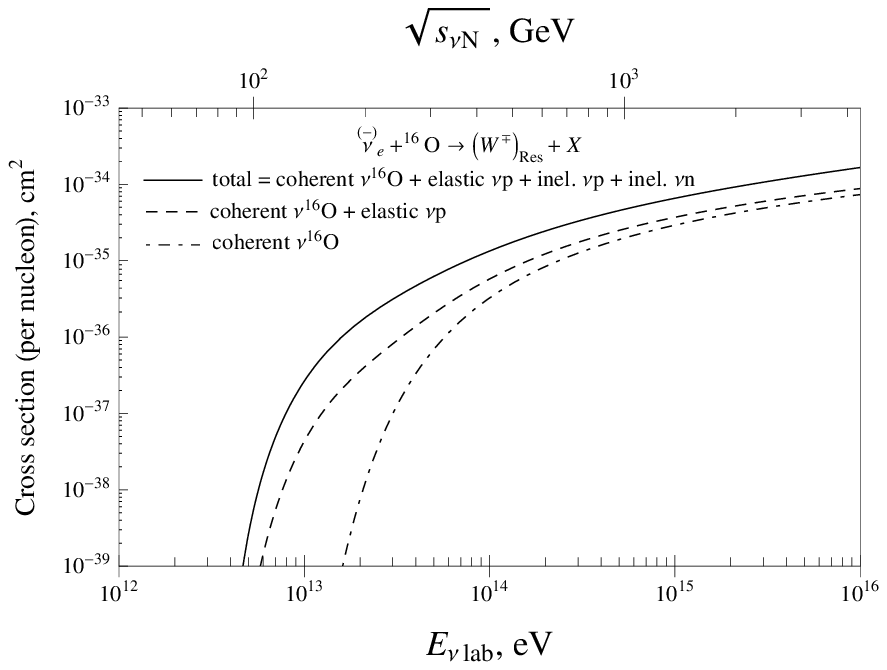}
\caption{}
\label{fig:cr_partial}
\end{figure}

In this section we present the cross sections for excitation of the Glashow resonance in neutrino scattering on the oxygen nucleus, $
\overset{\text{{\tiny (}}-\text{{\tiny )}}}{\nu_l}+\text{$^{16}$O}\rightarrow (W^\mp)_{\text{Res}}+X$,  calculated numerically in the equivalent photon approximation for the neutrino laboratory energies between $5\times10^{12}~\text{eV}$  and~$10^{16}~\text{eV}$.  In the evaluation of~(\ref{nucleus}) we set $\alpha(m_W^2)=1/128$, $G_F=1.16\times10^{-5}$ $\text{GeV}^{-2}$, $m_e=0.0005$~GeV, $m_{\mu}=0.1056$~GeV, $m_{\tau}=1.7768$~GeV, $m_W=80.4000$~GeV~\cite{pdg}, $Z=A/2=8$. The coherent photon content of the $\text{$^{16}$O}$ nucleus has already been found in~\cite{me_2015}, so we just borrow $\gamma_{\text{coherent}}(x)$ from that work. The procedure of theoretical computation of the elastic photon distribution of the proton, $\gamma_{\text{p\,el}}(x)$, as well as the inelastic ones for the nucleon, $\gamma_{\text{p,n\,inel}}(x)$, has been developed and  studied in detail~\cite{Kniehl:1990iv,Gluck:2002fi,Pisano:2005kz} which we  also adopt here.  Note that the functions $\gamma_{\text{p,n\,inel}}(x)$ are scale dependent and we fix the scale to be equal to the energy squared of the subprocesses $
\overset{\text{{\tiny (}}-\text{{\tiny )}}}{\nu_l}l^{\mp}\rightarrow W^{\mp}$, $Q^2=m_W^2$, as it is usually done in similar calculations~\cite{Pisano:2004xv}. All these equivalent photon distributions are shown in Fig.~\ref{fig:pdf}. Using them in~(\ref{nucleus}) we have obtained the cross sections for $\overset{\text{{\tiny (}}-\text{{\tiny )}}}{\nu_l}+\text{$^{16}$O}\rightarrow (W^\mp)_{\text{Res}}+X$  depicted in Fig.~\ref{fig:cr_total}. To see the contributions from each component of the photon content of $\text{$^{16}$O}$ to an overall cross section it is enough to consider just one case, for example $
\overset{\text{{\tiny (}}-\text{{\tiny )}}}{\nu_e}+\text{$^{16}$O}\rightarrow (W^\mp)_{\text{Res}}+X$, shown in Fig.~\ref{fig:cr_partial}. The proportions between these  contributions to the cross section for any of the remainder reactions, $
\overset{\text{{\tiny (}}-\text{{\tiny )}}}{\nu_\mu}+\text{$^{16}$O}\rightarrow (W^\mp)_{\text{Res}}+X$, $
\overset{\text{{\tiny (}}-\text{{\tiny )}}}{\nu_\tau}+\text{$^{16}$O}\rightarrow (W^\mp)_{\text{Res}}+X$, will be the same as above. It should be emphasized that the cross section for coherent $\overset{\text{{\tiny (}}-\text{{\tiny )}}}{\nu_l}\text{$^{16}$O}$ scattering is about two times lower than the result from~\cite{seckel}. This discrepancy is presumably caused by different treating the nuclear formfactor.

\begin{figure}
\includegraphics[width=1.\textwidth]{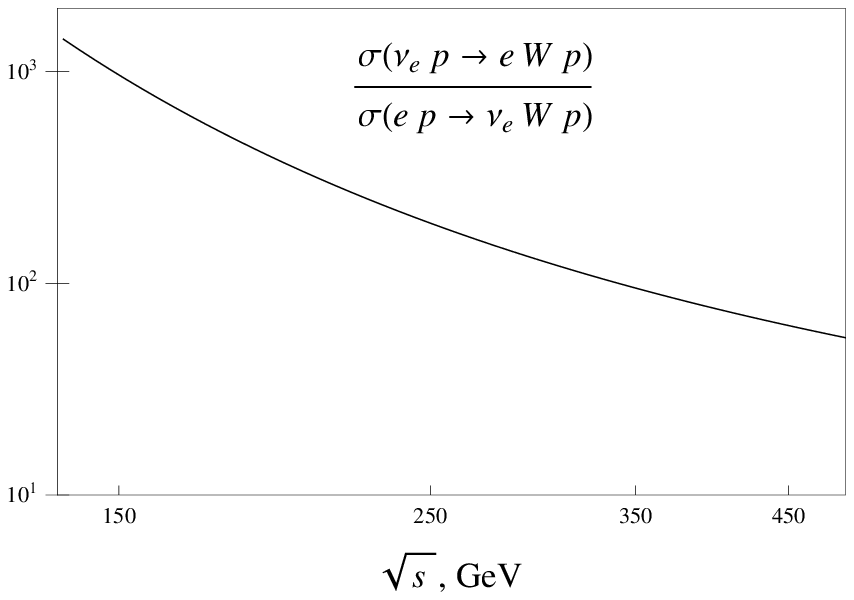}
\caption{}
\label{fig:ratio}
\end{figure}

A superficial look at the cross sections in Fig.~\ref{fig:cr_total} does not reveal the resonant behavior, but a slow logarithmic-law growth with the collision energy. The Glashow resonance is thus hidden in these reactions. 
Nevertheless, its presence can be seen already at the level of elastic scattering. As an example compare two processes, $\nu_ep\rightarrow e^-W^+p$  and $e^-p\rightarrow\nu_eW^-p$. Both processes proceed through the charged week current interaction and at $E_\nu=E_e\gg m_e$ are obviously similar to each other from the kinematical point of view. However there is a dramatic difference (by a factor of~$\sim100$) between their cross sections in a wide range of energies, as seen from Fig.~\ref{fig:ratio}. This difference can be related neither to the different available phase spaces ($e^-p\rightarrow\nu_eW^-p$ has a larger phase space compared to $\nu_ep\rightarrow e^-W^+p$) nor to averaging over the initial spins of the colliding particles (which gives only a factor of~2). This cannot also be due to large errors in the calculations because it has already been demonstrated that the equivalent photon approximation for such elastic processes reproduces the cross sections to a remarkable accuracy of below $1\%$~\cite{Kniehl:1990iv}. The reason for this difference is dynamical, namely the Glashow resonance. The matter is that while both reactions are dominated by photon exchange, the Standard Model forbids  direct splitting $\gamma\rightarrow\nu_e\bar\nu_e$ and therefore $e^-p\rightarrow\nu_eW^-p$ cannot involve the resonant subprocess $e^-\bar\nu_e\rightarrow W^-$ at~$O\left(\alpha G_F\right)$, when $\nu_ep\rightarrow e^-W^+p$ proceeds through $\nu_ee^+\rightarrow W^+$ due to the possibility $\gamma\rightarrow e^+e^-$. 

\section{Uncertainties and background}
The main source of uncertainties on the calculated cross sections is related to the uncertainty on the equivalent photon distribution $\gamma(x)$ in~(\ref{nucleus}).  The relative error of a cross section for an inelastic reaction as found in the framework of the equivalent photon approximation with respect to the exact result will depend on the four-momentum transfer squared $Q^2$ (the scale). 
It is essential that the photon distributions of the nucleon used above have already been theoretically tested for $W$ production in  $ep\rightarrow \nu_e WX$~\cite{Kniehl:1990iv,Pisano:2004xv} which is kinematically similar to the reactions we study in the sense that the scales at which all  these reactions proceed are obviously identical. Therefore our results reproduce the cross sections to the same accuracy as those in~\cite{Kniehl:1990iv,Pisano:2004xv}. Namely, at energies about $E_\nu=10^{13}~\text{eV}$ the uncertainties on the elastic and inelastic part of a cross section do not exceed $1\%$ and $10\%$, respectively. At higher energies, $E_\nu\sim10^{16}~\text{eV}$, they are less than $1\%$ and $3\%$, respectively. In total, one has that in the considered energy range the relative error for a cross section in Fig.~\ref{fig:cr_total} drops from $\sim5\%$ down to $\sim2\%$ as the neutrino energy increases. Thus, the equivalent photon approximation for the reactions $\overset{\text{{\tiny (}}-\text{{\tiny )}}}{\nu_l}+\text{$^{16}$O}\rightarrow (W^\mp)_{\text{Res}}+X$ and $\overset{\text{{\tiny (}}-\text{{\tiny )}}}{\nu_l}+p\rightarrow (W^\mp)_{\text{Res}}+X$ is quite satisfactory.

It should be noted, that the applicability of the equivalent photon approximation to the description of the reaction $ep\rightarrow \nu_e WX$ is only to be experimentally verified. Its contribution to the total cross section of single W boson production at HERA is about 7\%~\cite{Aaron:2009ab}, which is still at the level of measurement uncertainties. At the same time, experimental data on the deep inelastic Compton scattering, $ep\rightarrow e\gamma X$, whose cross section can also be expressed in terms of the equivalent photon distribution of the proton, convoluted with the real photoproduction cross section, $e\gamma\rightarrow e\gamma$~\cite{Gluck:1994vy,DeRujula:1998yq},  are accurately described by the equivalent photon approximation~\cite{Pisano:2005kz}.

\begin{figure}
\includegraphics[width=1.\textwidth]{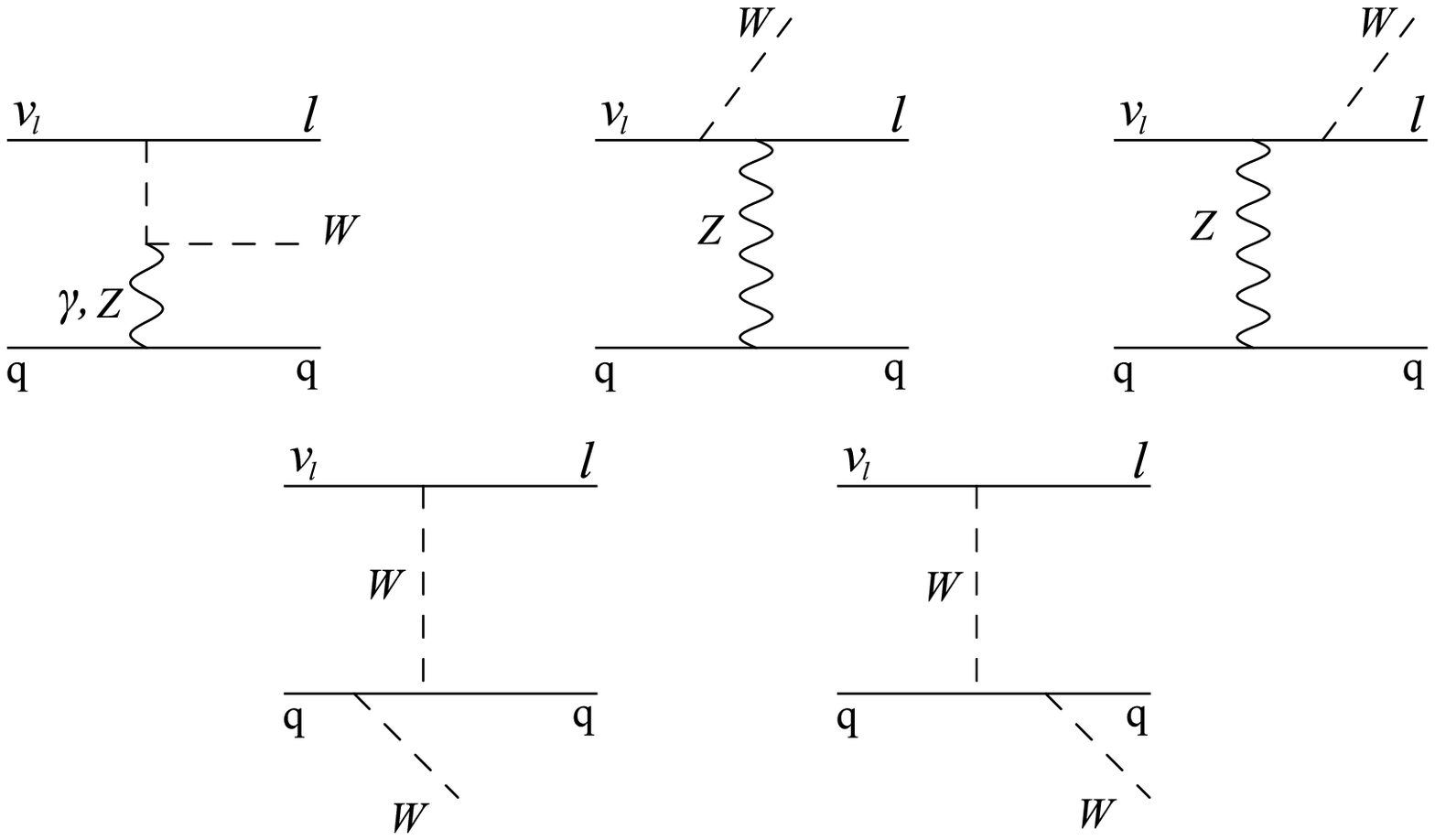}
\caption{}
\label{fig:back}
\end{figure}

There are other channels of~$W$ boson production in neutrino--nucleus collisions represented in the parton-level diagrams of Fig.~\ref{fig:back}.   These channels will constitute a background to searches for the Glashow resonance and one must know the related contribution to the overall event rate. To evaluate the background let us note that the lowest order diagrams for the reaction $ep\rightarrow \nu_e WX$ have exactly the same structure as 
those in~Fig.~\ref{fig:back}~\cite{Atwood:1990xg} (this is not surprising since we have already shown above that $W$ bosons in this case should emerge through non-resonant subprocesses).  This means that taking the ratio of the cross section for~$ep\rightarrow \nu_e WX$ to any of the cross sections  in Fig.~\ref{fig:cr_total} one automatically evaluates the relative contribution of the background to the $W$ boson production rate. Thus, for neutrino energies between~$10^{13}~\text{eV}$ and~$10^{14}~\text{eV}$, where the $\gamma$-exchange dominates, the contribution of the background reactions is below~$1\%$ and increases only to about~$2\%$ at~$E_\nu\sim10^{16}~\text{eV}$, where the $Z$-exchange diagrams become comparable in importance. An impressive view about the tiny role of the non-resonant channels in $W$ boson production compared to neutrino resonant scattering is also provided by~Fig.~\ref{fig:ratio} if to invert the ratio (in fact, the ratio~$\sigma(ep\rightarrow \nu_e W p)/\sigma(\nu_e p\rightarrow e W p)$ even overestimates this role due to the available phase space for $ep\rightarrow \nu_e Wp$ is larger than that for $\nu_e p\rightarrow e Wp$). These estimates hold not only for the case of $\overset{\text{{\tiny (}}-\text{{\tiny )}}}{\nu_e}$ scattering, but apparently for the reactions with $\overset{\text{{\tiny (}}-\text{{\tiny )}}}{\nu_\mu}$ and $\overset{\text{{\tiny (}}-\text{{\tiny )}}}{\nu_\tau}$ as well.

Summarizing the last paragraph we arrive at an important conclusion that with an uncertainty less than~$2\%$ all $W$  bosons produced in the sub-PeV region in neutrino-initiated reactions in water/ice  will be from the Glashow resonance. 

\section{Experimental observability}
By virtue of the wide variety of decay
modes, a $W$ boson produced through the Glashow resonance may have a rich set of possible signatures in a neutrino detector. Before discussing the expected signal, we estimate the Glashow resonance event rate per year
per km$^3$ water equivalent volume. Let us consider only the so-called downward-going events initiated by the neutrino flux from the upper hemisphere (to estimate the number of upward-going events one has to take into account the Earth attenuation effects (see, for example,~\cite{Chen:2013dza,Palomares-Ruiz:2015mka})). Thus, the event rate can be written as

\begin{equation}
N_W=2\pi TN_t\,\sum_{l=e,\mu,\tau}\int{dE_{\nu}\,\sigma_{Nl}(E_{\nu})\Phi_{\nu_l+\bar\nu_l}(E_{\nu})},\label{eq_distribe}
\end{equation}

where the integration is over a neutrino energy bin of interest, $N_t\simeq6\times10^{38}$ is the number of target nucleons in the volume, $T\simeq315\times10^5~\text{s}$ is the time of exposure, $\Phi_{\nu_l+\bar\nu_l}$~is the flux of neutrinos plus antineutrinos of flavor $l$. It is easy within our approach to take account the presence (apart from the $^{16}\text{O}$ nucleus) of two protons in each molecule of water. To do it one has just to put $Z=10$ and $A=18$ in~(\ref{eq:contribs}). 
The neutrino flux in~(\ref{eq_distribe}) is a superposition of the conventional neutrino flux, the prompt neutrino flux and the astrophysical neutrino flux: 

\begin{equation}
\Phi_{\nu_l+\bar\nu_l}=\Phi^{\text{conventional}}_{\nu_l+\bar\nu_l} + \Phi^{\text{prompt}}_{\nu_l+\bar\nu_l}+\Phi^{\text{astrophysical}}_{\nu_l+\bar\nu_l}. 
\end{equation}

For $\Phi^{\text{conventional}}_{\nu_l+\bar\nu_l}$ and $\Phi^{\text{prompt}}_{\nu_l+\bar\nu_l}$,  when $l=e,~\mu$, we adopt the corresponding parametrizations from~\cite{Sinegovskaya:2014pia} and set $\Phi^{\text{conventional}}_{\nu_\tau+\bar\nu_\tau}=\Phi^{\text{prompt}}_{\nu_\tau+\bar\nu_\tau}=0$. The astrophysical neutrinos comes into play at  $E_{\nu}\gtrsim100~\text{TeV}$ and in this region we take the best-fit flux $\Phi^{\text{astrophysical}}_{\nu_l+\bar\nu_l}\simeq0.95\times10^{-8}\left(E_{\nu}/\text{GeV}\right)^{-2}(\text{GeV}\,\text{cm}^{2}\,\text{s}\,\text{sr})^{-1}$ for each neutrino flavor~\cite{icecube_new}. The obtained numerical results are given in Tab.~\ref{tab:tab1}. 

First of all, one can see that the vast majority of the Glashow resonance events is expected at energies from a few TeV to a few tens of TeV, being mostly initiated by the conventional atmospheric neutrinos dominant in this energy range.   Since the Earth attenuation effects are relevant only at energies above 100~TeV, one can estimate the total number of the Glashow resonance events (upward-going~+~downward-going) for $E_{\nu}\lesssim~50~\text{TeV}$ just by doubling the corresponding  number of downward-going events from~Tab.~\ref{tab:tab1}. Thus, our prediction is~$\sim10$~Glashow resonance events per year per km$^3$ water equivalent volume at energies below~50~TeV, assuming the neutrino flux quoted above. 

High energy hadrons, electrons and tau leptons are usually visible at neutrino telescopes, such as IceCube, in the form of showers while muons give tracks.  It is apparent that the probability that a resonance event will manifest itself as a shower highly boosted along the incident neutrino path is given by 

\begin{equation}
\frac{\Gamma(W\rightarrow\text{hadrons})+\Gamma(W\rightarrow\nu_ee)+\Gamma(W\rightarrow\nu_\tau\tau)}{\Gamma(W\rightarrow\text{all})}\simeq0.9
\end{equation}

and the probability for observing a track is

\begin{equation}
\frac{\Gamma(W\rightarrow\nu_\mu\mu)}{\Gamma(W\rightarrow\text{all})}\simeq0.1,
\end{equation}

where $\Gamma(W\rightarrow\text{anything})$ is the width of the decay $W\rightarrow\text{anything}$.
Then, the track-to-shower ratio for the sample of the Glashow resonance events will be~$\sim0.1$ no matter what the flavor composition of the neutrino flux is. Thus, we can specify the above result: $\sim$~$9/\text{year\,km$^3$}$ Glashow resonance events with shower-like topologies and  $\sim1/\text{year\,km$^3$}$ tracks.  The deposited energies for both types of events will lie roughly between~$2.5$ and~$50$~TeV. 

The Glashow resonance at such low energies could be identified as follows. The W bosons will be excited mostly by atmospheric muon (anti)neutrinos, which dominate in this energy region, in the reactions  $\overset{\text{{\tiny (}}-\text{{\tiny )}}}{\nu_\mu}+\text{target}\rightarrow (W^\mp)_{\text{Res}}+\mu^{\pm}+X$. The subsequent very quick decays of the leading W bosons into hadrons, electrons and tau leptons will make significant contributions to the energies of the showers X, while the muons come from the target fragmentation causing thus the ratio $E_{\text{shower}}/E_{\text{track}}$ on average to be larger  than in  background events from the charged current neutrino scattering $\overset{\text{{\tiny (}}-\text{{\tiny )}}}{\nu_\mu}+\text{target}\rightarrow \mu^{\pm}+X$. The track of W would not be observed directly, but the boson could manifest itself from the lower-than-expected energy of the muon track. Due to event-to-event variations in the ratio, this analysis would have to be done
on a statistical basis.  In addition, such Glashow resonance events will be distributed anisotropically over the sky being concentrated mostly near the horizon as the incident conventional $\nu_{\mu}+\bar\nu_{\mu}$ flux does. 
This procedure is similar to that proposed for detecting low energy tau (anti)neutrinos in large volume Cherenkov detectors via the muonic tau decay~\cite{DeYoung:2006fg}.
It is worth to highlight that the Glashow resonance events can thus constitute a significant background to searches for the atmospheric $\nu_{\mu}+\bar\nu_{\mu}\rightarrow\nu_{\tau}+\bar\nu_{\tau}$ oscillations at $E_\nu>1~\text{TeV}$ if one uses the procedure of tau neutrino detection mentioned above. The tau-like event rate mimicked by the W excitations may be at least an order of magnitude higher than that evaluated, for example, by the authors of~\cite{Pasquali:1998xf}, depending on the adopted oscillation parameters.

It is also interesting to estimate the Glashow resonance event rate for~$E_{\nu}\gtrsim300~\text{TeV}$ expected at IceCube. In this energy range IceCube  has detected only~4 neutrino-initiated showers and no tracks for~988 days of observations~\cite{icecube_new} ({\it i.e.}, roughly~1.5 showers per year). 
To make our predictions applicable to IceCube observations we have to take into consideration the effective volume of the IceCube detector  (which, at these energies,~$\simeq0.4~\text{km}^3$~\cite{icecube_28}) as well as the attenuation factors for the components of the neutrino flux reaching the detector~\cite{Palomares-Ruiz:2015mka}. Thus, if we assume a~$40\%$~all-sky averaged neutrino flux attenuation (this presumably overestimates the actual attenuation effect), we obtain the total (upward-going~+~downward-going) Glashow resonance event rate in the IceCube detector  in this energy region to be~$\sim0.3$ per year.  Accordingly, it is likely to have at least one Glashow resonance event with a shower-like topology and the deposited energy~$\gtrsim300~\text{TeV}$ in the IceCube data set (the data taking time already is about~3 years).

\section{Conclusions}
Today it is widely believed that $s$-channel excitation of an on-shell~$W$ boson, commonly known as the Glashow resonance, can be initiated in matter only by the electron antineutrino in the process $\bar\nu_ee^-\rightarrow W^-$  at the laboratory energy  around~6.3~PeV.
In this Letter we argue that the Glashow resonance within the Standard Model also occurs in neutrino--nucleus collisions. 

Our conclusions are as follows.

1)~The Glashow resonance can be excited  by both~neutrinos and~antineutrinos of all the three flavors scattering in the Coulomb field of a nucleus.

2)~The Glashow resonance in a neutrino--nucleus reaction does not manifest itself as a Breit--Wigner-like peak in the cross section but the latter exhibits instead a slow logarithmic-law growth with the neutrino energy. The resonance turns thus out to be hidden.  

3)~More than~$98\%$ of~$W$  bosons produced in the sub-PeV region in neutrino-initiated reactions in water/ice  will be from the Glashow resonance. 

4)~The vast majority of the Glashow resonance events in a neutrino detector is expected at energies from a few TeV to a few tens of TeV, being mostly initiated by the conventional atmospheric neutrinos dominant in this energy region. It is explained how the resonance events could be identified at such low energies.

5)~About~90\% of the Glashow resonance events in a  given data sample will be in the form of showers boosted along the incident neutrino path and only $\sim10\%$ as tracks.

Calculations of the cross sections for Glashow resonance excitation  on the oxygen nucleus as well as on the proton are carried out in detail. 
The results of this Letter can be useful for studies of neutrino interactions at large volume neutrino detectors as the IceCube  detector~\cite{icecube_rev},  the ANTARES undersea neutrino telescope~\cite{antares} as well as the next generation deep-water neutrino telescopes KM3NeT~\cite{km3net} and NT1000 on Lake Baikal~\cite{baikal}. 
For example, in the IceCube detector one can expect~0.3 Glashow resonance events  with  shower-like topologies and the deposited energies above~$300~\text{TeV}$ per year. It is therefore likely to have at least one such resonance event in the IceCube data set (the data taking time already is about~3 years).

The theoretical framework of this Letter is readily applicable to description of neutrino resonant scattering on different nuclear targets~\cite{me_baldin}. It is also fair to expect that other $s$-channel neutrino-initiated reactions, such as $\overset{\text{{\tiny (}}-\text{{\tiny )}}}{\nu_e}e^{\mp}\rightarrow\rho^{\mp}$~\cite{Paschos:2002sj}, can be experimentally tested in neutrino--nucleus collisions.

\vskip 0.5cm
{\bf Acknowledgements}
\vskip 0.5cm

I wish to thank E.~A.~Paschos, S.~I.~Sinegovsky and A.~C.~Vincent for
valuable comments. I am also grateful to H.~P\"as for kindly inviting
me to attend weekly meetings of his research group at which I
have had a nice opportunity to discuss this work. This work was
supported in part by the Program for Basic Research of the Presidium
of the Russian Academy of Sciences “Fundamental Properties
of Matter and Astrophysics”. I acknowledge DAAD support through
the funding program “Research Stays for University Academics and
Scientists” and hospitality at TU Dortmund where this work was
finished.


\newpage

{\bf Figure captions}
\vskip 0.5 cm 

{\bf Fig. 1:} Total cross sections for $\overset{\text{{\tiny (}}-\text{{\tiny )}}}{\nu_l}\gamma\rightarrow l^\pm W^\mp$ $(l=e,\mu,\tau)$ as functions of the ratio $s/m_W^2$.

\vskip 0.5 cm

{\bf Fig. 2:} A schematic illustration of the initial state lepton emission mechanism for excitation of the Glashow resonance in ${\nu_l}\gamma\rightarrow W^+l^-$ $(l=e, \mu, \tau)$. The photon  splits into an $l^+l^-$ pair  before the excitation occurs. Even if the center-of-mass energy of the $\nu_l\gamma$ collision, $\sqrt{s}$, exceeds the $W$ boson mass, $m_W$, the emitted lepton $l^-$ carries away the energy excess $E=\sqrt{s}-m_W$ and turns thus the energy of the $\nu_ll^+$ pair to the resonance pole.

\vskip 0.5 cm

{\bf Fig. 3:} Different sources of the equivalent (Weizs\"acker--Williams) photons with which neutrinos interact: (a)~the coherent photon content of a nucleus; (b)~elastic photon content of the proton;  (c)~inelastic photon content of the nucleon (proton and neutron).

\vskip 0.5 cm

{\bf Fig. 4:} The equivalent (Weizs\"acker--Williams) photon distributions~\cite{me_2015,Gluck:2002fi}. The inelastic photon distributions for the proton and neutron are taken at a fixed scale~$Q^2=m_W^2$.

\vskip 0.5 cm

{\bf Fig. 5:} Per nucleon total cross sections for~$\overset{\text{{\tiny (}}-\text{{\tiny )}}}{\nu_l}+\text{$^{16}$O}\rightarrow (W^\mp)_{\text{Res}}+X$  $(l=e,\mu,\tau)$ as functions of the neutrino laboratory energy. The corresponding center-of-mass  neutrino--nucleon collision energy  is labeled  on the upper horizontal axis.

\vskip 0.5 cm

{\bf Fig. 6:} Different parts of the per nucleon total cross section for $\overset{\text{{\tiny (}}-\text{{\tiny )}}}{\nu_e}+\text{$^{16}$O}\rightarrow (W^\mp)_{\text{Res}}+X$ as functions of the neutrino laboratory energy. The corresponding center-of-mass  neutrino--nucleon collision energy  is labeled  on the upper horizontal axis.

\vskip 0.5 cm

{\bf Fig. 7:} Ratio of the cross section for $\nu_ep\rightarrow eWp$ (calculated in this Letter) to that for $ep\rightarrow \nu_eWp$~\cite{Kniehl:1990iv} as a function of the center-of-mass collision energy.

\vskip 0.5 cm

{\bf Fig. 8:} Diagrams for the parton level processes contributing to the background. 


\newpage

\begin{table}
\caption{The Glashow resonance event rate per year
per km$^3$ water equivalent volume. Only the downward-going events are presented.
}
\vskip 0.3cm
\begin{tabular}{lccc}
\hline
\hline
Neutrino energy \,\,\,\,\,& \,\,\,\,\,$5-50~\text{TeV}$ \,\,\,\,\,& \,\,\,\,\,$50-300~\text{TeV }$ \,\,\,\,\,&\,\,\,\,\, $>300~\text{TeV}$\,\,\,\,\, 
 \\ \hline
$N_W$ \,\,\,\,\,& \,\,\,\,\,5.2 \,\,\,\,\,& \,\,\,\,\,2.3\,\,\,\,\,& \,\,\,\,\,0.6\,\,\,\,\,\\
\hline
\hline
\end{tabular}
\label{tab:tab1}
\end{table}

\end{document}